\documentclass[technote]{IEEEtran} 
\usepackage{graphicx}
\usepackage{dcolumn}
\usepackage{array}
\usepackage{amsmath}
\usepackage{fancyheadings}
\usepackage{subfigure}
\usepackage{algorithm}
\usepackage{algorithmic}
\usepackage{verbatim}
\newcolumntype{d}[1]{D{.}{.}{#1}}
\newcolumntype{i}{D{.}{}{0}}

\begin{document}
\pagestyle{empty}
\title{Optimal Transmission Strategy and Explicit Capacity Region for Broadcast Z Channels}
\author{\normalsize{Bike Xie, {\it Student Member,
IEEE}, Miguel Griot, {\it Student Member, IEEE}, Andres I. Vila
Casado, {\it Student Member, IEEE} and Richard D. Wesel, {\it Senior
Member, IEEE}}
\thanks{This work was supported by the Defence Advanced Research
Project Agency SPAWAR Systems Center, San Diego, California under
Grant N66001-02-1-8938. This paper was presented in part at the
Information Theory Workshop 2007.
\newline The authors are with the Electrical Engineering Department,
University of California, Los Angeles, CA 90095 USA
(e-mail:xbk@ee.ucla.edu; mgriot@ee.ucla.edu; avila@ee.ucla.edu;
wesel@ee.ucla.edu).}}

\maketitle \thispagestyle{plain}
\begin {abstract}
This paper provides an explicit expression for the capacity region
of the two-user broadcast Z channel and proves that the optimal
boundary can be achieved by independent encoding of each user.
Specifically, the information messages corresponding to each user
are encoded independently and the OR of these two encoded streams is
transmitted. Nonlinear turbo codes that provide a controlled
distribution of ones and zeros are used to demonstrate a
low-complexity scheme that operates close to the optimal boundary.
\end {abstract}
\begin {keywords}
broadcast channel, broadcast Z channel, capacity region, nonlinear
turbo codes, turbo codes.
\end {keywords}

\section{Introduction}

Degraded broadcast channels were first studied by Cover in
\cite{Cover1972} and a formulation of the capacity region was
established in \cite{Bergmans1973}, \cite{Bergmans1974} and
\cite{Gallager1974}. Superposition encoding is the key idea to
achieve the optimal boundary of the capacity region for degraded
broadcast channels \cite{Cover1998}. With superposition encoding for
degraded broadcast channels, the data sent to the user with the most
degraded channel is encoded first. Given the encoded bits for that
user, an appropriate codebook for the second most degraded channel
user is selected, and so forth. Hence superposition encoding is, in
general, a joint encoding scheme. However, combining independently
encoded streams, one for each user, is an optimal scheme for some
broadcast channels including broadcast Gaussian channels
\cite{Cover1972} and broadcast binary-symmetric channels
\cite{Cover1972} \cite{Bergmans1973}.

Successive decoding is a natural decoding scheme for superposition
encoding \cite{Cover1972} \cite{Bergmans1973} \cite{Cover1998}. With
successive decoding for degraded broadcast channels, each receiver
first decodes the data sent to the user with the most degraded
channel. Conditioning on the decoded data for that user, each
receiver determines the codebook for the user with the second most
degraded channel and decodes that data, and so forth until the
desired user's data is decoded. The performance of successive
decoding for degraded broadcast channels is very close to optimal
decoding under normal operating conditions.

Turbo codes \cite{Berrou1993} and Low-Density Parity-Check (LDPC)
codes \cite{Gallager1963} perform close to the Shannon limit. LDPC
and turbo coding approach for broadcast channels were studied in
\cite{Berlin2004} and \cite{ThomasSun2004} respectively. In
\cite{Berlin2004}, LDPC codes provided reliable transmission over
two-user broadcast channels with additive white Gaussian noise
(AWGN) and fading known at the receiver only. In
\cite{ThomasSun2004},  a superposition turbo coding scheme performs
within 1dB of the capacity region boundary for broadcast Gaussian
channels. Both of these approaches are designed specifically for
broadcast Gaussian channels and used linear codes. For multi-user
binary adder channels, nonlinear trellis codes were studied and
designed in \cite{Chevillat1981}.

\begin{figure}
  \centering
  \includegraphics[width=0.4\textwidth]{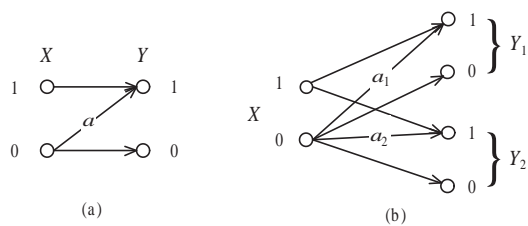}
  \caption{(a) Z channel. (b) Broadcast Z channel.}\label{fig:BZChan}\vspace{-0.1in}
\end{figure}

The Z channel is the binary-asymmetric channel shown in
Fig.~\ref{fig:BZChan}(a). The capacity of the Z channel was studied
in \cite{Golomb1980}. Nonlinear trellis codes were designed to
maintain a low ones density for the Z channel in \cite{griot-ISIT06}
\cite{griot-TCOM} and parallel concatenated nonlinear turbo codes
were designed for the Z channel in \cite{griot-GC06}. This paper
focuses on the study of the two-user broadcast Z channel $X
\rightarrow Y_1,Y_2$ shown in Fig.~\ref{fig:BZChan}(b). This paper
provides an explicit expression of the capacity region for the
two-user broadcast Z channel and shows that independent encoding
with successive decoding can achieve the boundary of this capacity
region.

This paper is organized as follows. Section \ref{sec:BZC} introduces
 definitions and notation for broadcast channels. Section \ref{sec:ots} provides the explicit expression of the capacity
region for the two-user broadcast Z channel and proves that
independent encoding can achieve the optimal boundary of the
capacity region. Section \ref{sec:nlturbo} presents nonlinear-turbo
codes designed to achieve the optimal boundary, and Section
\ref{sec:results} provides the simulation results. Section
\ref{sec:conclusions} delivers the conclusions.

\section{Definitions and Preliminaries}
\label{sec:BZC}

\subsection{Degraded broadcast channels}

\begin{figure}
  \centering
  \includegraphics[width=0.45\textwidth]{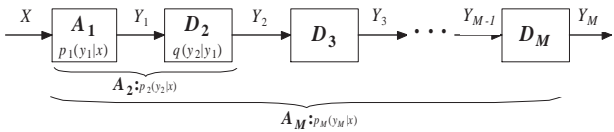}
  \caption{Physically degraded broadcast channel.}\label{fig:DBC}\vspace{-0.1in}
\end{figure}

The general representation of a discrete memoryless broadcast
channel is given in Fig.~\ref{fig:BC}. A single signal $X$ is
broadcast to $M$ users through $M$ different channels $A_1, \cdots,
A_M$. If $p ( y_i , y_{i+1} | x ) = p ( y_i | x ) p ( y_{i+1} | y_i
)$, then channel $A_{i+1}$ is a physically degraded version of
channel $A_{i}$ (and thus the broadcast channel $X\rightarrow Y_{i},
Y_{i+1}$ is physically degraded) \cite{Cover1998}. A physically
degraded broadcast channel with $M$ users is shown in
Fig.~\ref{fig:DBC}. Since each user decodes its received signal
without collaboration, only the marginal transition probabilities
$p(y_{1}|x), p(y_{2}|x),\cdots ,p(y_{M}|x)$ of the component
channels $A_{1}, A_{2},\cdots,A_{M}$ affect receiver performance.
Hence, the \textit{stochastically} degraded broadcast channel is
defined in \cite{Bergmans1973} and \cite{Cover1998} as follows:

\begin{figure}
  \centering
  \includegraphics[width=0.24\textwidth]{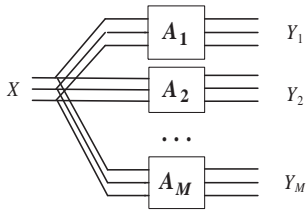}
  \caption{Broadcast channel.}\label{fig:BC}\vspace{-0.1in}
\end{figure}

Let $A_{i}$ be a channel with input alphabet $\mathcal{X}$, output
alphabet $\mathcal{Y}_{\mathrm{i}}$, and transition probability
$p_{i}(y_{i}|x)$. Let $A_{i+1}$ be another channel with the same
input alphabet $\mathcal{X}$, output alphabet
$\mathcal{Y}_{\mathrm{i+1}}$, and transition probability
$p_{i+1}(y_{i+1}|x)$. $A_{i+1}$ is a stochastically degraded version
of $A_{i}$ if there exists a transition probability
$q(y_{i+1}|y_{i})$ such that
\begin{equation}
p_{i+1}(y_{i+1}|x)=\sum_{y_{i}\in
\mathcal{Y}_{\mathrm{i}}}{q(y_{i+1}|y_{i})p_{i}(y_{i}|x)}.
\end{equation}

A broadcast channel with receivers $Y_{1},Y_{2}\cdots ,Y_{M}$ is a
stochastically degraded broadcast channel if every component channel
$A_{i}$ is a stochastically degraded version of $A_{i-1}$ for all
$i=2,\cdots,M$ \cite{Bergmans1973}. Since the marginal transition
probabilities $p(y_{1}|x), p(y_{2}|x),\cdots ,p(y_{M}|x)$ completely
determine a stochastically degraded broadcast channel, we can model
any stochastically degraded broadcast channel as a physically
degraded broadcast channel with the same marginal transition
probabilities.

\newtheorem{degradedBC}{Theorem}
\begin{degradedBC}[\cite{Bergmans1973}
\cite{Gallager1974}]\label{theorem:capaDBC} The capacity region for
the two-user stochastically degraded broadcast channel $X \to Y_{1}
\to Y_{2}$ is the convex hull of the closure of all $(R_{1}, R_{2})$
satisfying
\begin{equation}
R_{2} \leq I(X_2;Y_{2}) \qquad R_{1}\leq I(X;Y_{1}|X_2),
\end{equation}
for some joint distribution $p(x_2)p(x|x_2)p(y_1,y_2|x)$, where the
auxiliary random variable $X_2$\footnote{$U$ was used as the
auxiliary random variable in \cite{Bergmans1973}
\cite{Gallager1974}. In this paper, we use $X_2$ instead of $U$
because the auxiliary random variable corresponds to the second
user's encoded stream.} has cardinality bounded by
$|\mathcal{X}_{\mathrm{2}}|\leq\min{\{|\mathcal{X}|,|\mathcal{Y}_{\mathrm{1}}|,|\mathcal{Y}_{\mathrm{2}}|\}}$.
\end{degradedBC}

\subsection{The broadcast Z channel}

\begin{figure}
  \centering
  \includegraphics[width=0.24\textwidth]{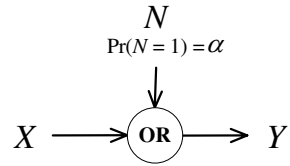}
  \caption{OR operation view of Z channel.}\label{fig:ZOR}\vspace{-0.1in}
\end{figure}

\begin{figure}
  \centering
  \includegraphics[width=0.35\textwidth]{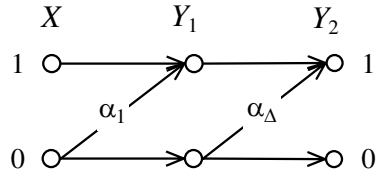}
  \caption{Physically degraded broadcast Z channel.}\label{fig:DBZC}\vspace{-0.1in}
\end{figure}

The Z channel, shown in Fig.~\ref{fig:BZChan}(a), is a
binary-asymmetric channel with the transition probability matrix
\begin{equation} \label{eq:BBSC}
T=\begin{bmatrix}
1 & \alpha\\
0 & 1- \alpha
\end{bmatrix}, \nonumber
\end{equation}
where $0 \leq \alpha \leq 1$. If symbol 1 is transmitted, symbol 1
is received with probability 1. If symbol 0 is transmitted, symbol 1
is received with probability $\alpha$ and symbol 0 is received with
probability $1-\alpha$. We can model the Z channel as the OR
operation of the channel input $X$ and Bernoulli noise $N$ with
parameter $\alpha$ as shown in Fig.~\ref{fig:ZOR}. In an OR Multiple
Access Channel, each user appears to transmit over a Z channel when
the other users are treated as noise \cite{griot-GC06}. Thus, in an
OR network with multiple transmitters and multiple receivers, each
transmitter transmitting to more than one receiver sees a broadcast
Z channel if other transmitters transmitting to those receivers are
treated as noise. The two-user broadcast Z channel with the marginal
transition probability matrices
\begin{equation} \label{eq:BBSC}
T_1=\begin{bmatrix}
1 & \alpha_1\\
0 & 1- \alpha_1
\end{bmatrix} \quad \quad
T_2=\begin{bmatrix}
1 & \alpha_2\\
0 & 1- \alpha_2
\end{bmatrix} \nonumber
\end{equation}
is shown in Fig.~\ref{fig:BZChan}, where $0 \leq \alpha_1 \leq
\alpha_2 \leq 1$. Because broadcast Z channels are stochastically
degraded, we can model any broadcast Z channel as a physically
degraded broadcast Z channel as shown in Fig.~\ref{fig:DBZC}, where
\begin{equation}
\alpha_\Delta=\frac{\alpha_2-\alpha_1}{1-\alpha_1}.
\end{equation}

\section{Optimal Transmission Strategy for the Two-User Broadcast Z Channel}
\label{sec:ots}

\begin{figure}
  \centering
  \includegraphics[width=0.4\textwidth]{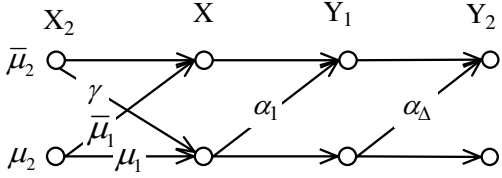}
  \caption{Information theoretic diagram of the system.}\label{fig:capBZC}\vspace{-0.1in}
\end{figure}

Since the broadcast Z channel is stochastically degraded, its
capacity region can be obtained directly from Theorem
\ref{theorem:capaDBC}. The capacity region for the broadcast Z
channel $X \to Y_{1} \to Y_{2}$ as shown in Fig.~\ref{fig:capBZC} is
the convex hull of the closure of all $(R_{1}, R_{2})$ satisfying
\begin{align}
\label{eq:capI2} R_{2} & \leq  I_2 = I(X_2;Y_{2}) \nonumber\\
& =  H\big((\bar{\mu}_2 \gamma  + \mu_2\mu_1)(1 - \alpha_2 )\big)
\nonumber\\
& \quad - \bar{\mu}_2H\big(\gamma (1 - \alpha_2 )\big) -
\mu_2H\big(\mu_1(1 - \alpha_2)\big),\\
\label{eq:capI1} R_{1} & \leq I_1 = I(X;Y_{1}|X_2) \nonumber\\
& = \bar{\mu}_2 \big(H(\gamma (1 - \alpha_1 )) - \gamma H(1 -
\alpha_1 )\big) \nonumber\\
& \quad + \mu_2\big(H(\mu_1(1 - \alpha_1 )) - \mu_1H(1 - \alpha_1
)\big),
\end{align}
for some probabilities $ \mu_1 , \mu_2, \gamma$, where
$\mu_1=\text{Pr}(x=0|x_2=0)$, $\mu_2=\text{Pr}(x_2=0)$, $\gamma
=\text{Pr}(x=0|x_2=1)$, $H(\cdot)$ is the binary entropy function,
$\bar{\mu}_1=1-\mu_1$, $\bar{\mu}_2=1-\mu_2$ and
\begin{equation}
\alpha_2=\text{Pr}\{y_2=1|x=0\}=1-(1-\alpha_1)(1-\alpha_\Delta).
\end{equation}

Each particular choice of $(\mu_1,\mu_2,\gamma)$ in
Fig.~\ref{fig:capBZC} specifies a particular transmission strategy
and a rate pair $(I_1,I_2)$. The optimal boundary of a capacity
region is the set of all Pareto optimal points $(I_1,I_2)$, for
which it is impossible to increase rate $I_1$ without decreasing
rate $I_2$ or vice versa. A transmission strategy is optimal if and
only if it achieves a rate pair point on the optimal boundary. We
call a set of transmission strategies \emph{sufficient} if all rate
pairs on the optimal boundary can be achieved by using these
strategies and time sharing. Furthermore, a set of transmission
strategies is \emph{strongly sufficient} if these strategies can
achieve all rate pairs on the optimal boundary without using time
sharing. Equations (\ref{eq:capI2}) and (\ref{eq:capI1}) give a set
of pentagons that yield the capacity region through their convex
hull, but do \emph{not} explicitly show the optimal transmission
strategies or derive the boundary of the capacity region.

\subsection{Optimal transmission strategies}

The following theorem identifies a set of optimal transmission
strategies and provides an explicit expression of the boundary of
the capacity region.

\newtheorem{strongopt}[degradedBC]{Theorem}
\begin{strongopt}\label{theorem:strongopt}
For a broadcast Z channel with $0<\alpha_1<\alpha_2<1$, the set of
the optimal transmission strategies $(\mu_1,\mu_2,\gamma)$, which
satisfy
\begin{equation}
\label{eq:ots_gamma}\gamma  = 0,
\end{equation}
\begin{equation}
\label{eq:ots_mu1}
\frac{1}{(1-\alpha_1)(e^{H(1-\alpha_1)/(1-\alpha_1)}+1)} \leq \mu_1
\leq 1,
\end{equation}
and
\begin{align}
\label{eq:ots_mu2} & \quad
\big(H(\mu_1(1-\alpha_1))-\mu_1H(1-\alpha_1)\big) \cdot
\ln(1-\mu_1(1-\alpha_2))
\nonumber \\
&=\big(H(\mu_1(1-\alpha_2)) -
\mu_1(1-\alpha_2)\ln\frac{1-\mu_2\mu_1(1-\alpha_2)}{\mu_2\mu_1(1-\alpha_2)}\big) \nonumber\\
& \quad \cdot \ln(1-\mu_1(1-\alpha_1)),
\end{align}
are strongly sufficient. In other words, all rate pairs on the
optimal boundary of the capacity region can be achieved by using
exactly the transmission strategies described in
(\ref{eq:ots_gamma}-\ref{eq:ots_mu2}) without the need of time
sharing. Furthermore, applying (\ref{eq:ots_gamma}-\ref{eq:ots_mu2})
to (\ref{eq:capI2}) and (\ref{eq:capI1}) yields an  explicit
expression of the optimal boundary of the capacity region.
\end{strongopt}

Before proving Theorem \ref{theorem:strongopt}, we present and prove
some preliminary results. From (\ref{eq:capI2}) and
(\ref{eq:capI1}), we can see that the transmission strategies
$(\mu_1,\mu_2,\gamma)$ and
 $(\gamma,1-\mu_2,\mu_1)$ have the same
transmission rate pairs. Therefore, we assume $\gamma\leq \mu_1$ in
the rest of the section without loss of generality.

\newtheorem{notopt}[degradedBC]{Theorem}
\begin{notopt}\label{theorem:notopt}
For a broadcast Z channel with $0<\alpha_1<\alpha_2<1$, any
transmission strategy $(\mu_1,\mu_2,\gamma)$ with $0< \mu_2<1,0<
\gamma< \mu_1$ is not optimal.
\end{notopt}

The proof is given in Appendix A.

\newtheorem{IESopt}{Corollary}
\begin{IESopt}\label{theorem:IESopt}
The set of all the transmission strategies with $\gamma = 0$ is
sufficient for any broadcast Z channel with $0<\alpha_1<\alpha_2<1$.
\end{IESopt}

\begin{figure}
  \centering
  \includegraphics[width=0.45\textwidth]{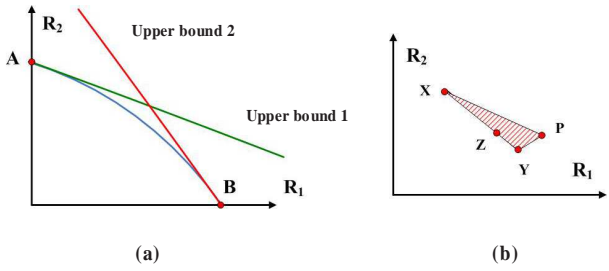}
  \caption{(a) The capacity region and two upper bounds. (b) Point $Z$ cannot be on the boundary of the capacity region.}\label{fig:detailcap}\vspace{-0.1in}
\end{figure}

\emph{Proof:} From Theorem \ref{theorem:notopt}, we know that the
transmission strategy $(\mu_1,\mu_2,\gamma)$ is optimal only if at
least one of these four equations $\mu_2=0$, $\mu_2=1$,
$\gamma=\mu_1$, $\gamma=0$ is true. Hence the set of all the
transmission strategies with $\mu_2=0$, $\mu_2=1$, $\gamma=\mu_1$ or
$\gamma=0$ is sufficient. When $\mu_2=0$, $\mu_2=1$ or
$\gamma=\mu_1$, the transmission rate for the second user, $I_2$ in
equation (\ref{eq:capI2}), is zero. This optimal rate pair is the
point $B$ in Fig.~\ref{fig:detailcap}(a). Since this point can also
be achieved by the transmission strategy with $\gamma=0$, $\mu_2=1$
and $\mu_1=\arg\max(H(x(1-\alpha_1))-xH(1-\alpha_1))$, all the
optimal rate pairs on the optimal boundary of the capacity region
can be achieved by using the transmission strategies with $\gamma=0$
and time sharing. Thus, the set of all the transmission strategies
with $\gamma = 0$ is sufficient. Q.E.D.

From Corollary \ref{theorem:IESopt}, we can set $\gamma = 0$ in
Fig.~\ref{fig:capBZC} without losing any part of the capacity region
and so the designed virtual channel $X_2 \rightarrow X$ is a Z
channel. Since we can consider the output of a Z channel as the OR
operation of two Bernoulli random variables, an independent encoding
scheme that works well for the broadcast Z channel will be
introduced later in this paper.

Applying $\gamma = 0$ to (\ref{eq:capI2}) and (\ref{eq:capI1})
yields
\begin{align}\label{eq:achI2}
R_{2} &\leq I_2 = H(\mu_2\mu_1(1 - \alpha_2 ))- \mu_2 H(\mu_1(1 -
\alpha_2
)),\\
\label{eq:achI1} R_{1} &\leq I_1 = \mu_2 H(\mu_1(1 - \alpha_1 )) -
\mu_2\mu_1 H(1 - \alpha_1 ).
\end{align}
By Corollary \ref{theorem:IESopt}, the capacity region is the convex
hull of the closure of all rate pairs $(R_1,R_2)$ satisfying
(\ref{eq:achI2}) and (\ref{eq:achI1}) for some probability
$\mu_1,\mu_2$. However, \emph{not} all transmission strategies of
$(\mu_1, \mu_2, \gamma = 0)$ achieve the optimal boundary of the
capacity region. Since any optimal transmission strategy maximizes
$I_1 + \lambda I_2$ for some nonnegative $\lambda$, we solve the
optimization problem of maximizing $I_1 + \lambda I_2$ for any fixed
$\lambda \geq 0$ in order to find the constraints on $\mu_1$ and
$\mu_2$ for optimal transmission strategies. Theorem
\ref{theorem:optdetail} provides the solution to this maximization
problem.

\newtheorem{optdetail}[degradedBC]{Theorem}
\begin{optdetail}\label{theorem:optdetail}
The optimal solution to the maximization problem
\begin{align}
 \textrm{maximize} \qquad &  I_1+\lambda I_2\label{eq:max}\\
 \textrm{subject to} \qquad & I_2=H(\mu_2\mu_1(1 - \alpha_2 ))- \mu_2
H(\mu_1(1 - \alpha_2 ))\nonumber\\
&I_1=\mu_2 H(\mu_1(1 - \alpha_1 )) - \mu_2\mu_1 H(1 - \alpha_1 )\nonumber\\
&0\leq \mu_2\leq1,0\leq \mu_1\leq1,\nonumber
\end{align}
is unique and it is given below for any fixed $\lambda \geq 0$.
\\Define
\begin{equation}
\varphi(x)=\frac{\ln(1-(1-\alpha_1)x)}{\ln(1-(1-\alpha_2)x)}
\end{equation}
and
\begin{equation}
\psi(x)=\frac{1}{xe^{H(x)/x}+x}.
\end{equation}
\emph{Case 1:} if $0 \leq\lambda \leq \varphi(\psi(1-\alpha_1))$,
then the optimal solution is $\mu_2^*=1,\mu_1^*=\psi(1-\alpha_1)$,
which satisfies (\ref{eq:ots_mu1}) and (\ref{eq:ots_mu2}), and the
corresponding rate pair is $I_1^*=H(\mu_1^*(1-\alpha_1))-\mu_1^*H(1-\alpha_1)$, $I_2^*=0$.\\
\emph{Case 2:} if $\lambda \geq \varphi(1)$, then the optimal
solution is $\mu_2^*=\psi(1-\alpha_2),\mu_1^*=1$, which also
satisfies (\ref{eq:ots_mu1}) and (\ref{eq:ots_mu2}), and the
corresponding rate pair is $I_1^*=0,I_2^*=H(\mu_2^*(1-\alpha_2))-\mu_2^*H(1-\alpha_2)$.\\
\emph{Case 3:} if $\varphi(\psi(1-\alpha_1)) <\lambda< \varphi(1)$,
then the optimal solution given below also satisfies
(\ref{eq:ots_mu1}) and (\ref{eq:ots_mu2}):
\begin{equation}\label{eq:opt_t01}
\mu_1^*=\varphi^{-1}(\lambda)=\frac{e^\lambda - 1}{e^\lambda
(1-\alpha_2)-(1-\alpha_1)}
\end{equation}
and
\begin{align}\label{eq:opt_pu}
& \quad \big(H(\mu_1^*(1-\alpha_1))-\mu_1^*H(1-\alpha_1)\big) \cdot
\ln(1-\mu_1^*(1-\alpha_2)) \nonumber \\
& = \big(H(\mu_1^*(1-\alpha_2)) -
\mu_1^*(1-\alpha_2)\ln\frac{1-\mu_2^*\mu_1^*(1-\alpha_2)}{\mu_2^*\mu_1^*(1-\alpha_2)}\big) \nonumber\\
& \quad \cdot \ln(1-\mu_1^*(1-\alpha_1)).
\end{align}

\end{optdetail}

The proof is given in Appendix B. Combining Case 1,2 and 3, we
conclude that $(\mu_1,\mu_2)$ is a maximizer of (\ref{eq:max}) if
and only if the pair $(\mu_1,\mu_2)$ satisfies (\ref{eq:ots_mu1})
and (\ref{eq:ots_mu2}). In other words, if $(\mu_1,\mu_2)$ doesn't
satisfy (\ref{eq:ots_mu1}) or (\ref{eq:ots_mu2}), $(\mu_1,\mu_2)$
cannot be a maximizer of (\ref{eq:max}), and thus the transmission
strategy $(\mu_1,\mu_2,\gamma =0)$ is not optimal. Since the set of
the transmission strategies with $\gamma = 0$ is sufficient by
Corollary \ref{theorem:IESopt}, the set of all the transmission
strategies satisfying (\ref{eq:ots_gamma}-\ref{eq:ots_mu2}) is also
sufficient. Therefore the capacity region is the convex hull of the
closure of all rate pairs $(R_1,R_2)$ satisfying (\ref{eq:achI2})
and (\ref{eq:achI1}) for some $\mu_1,\mu_2$ which satisfy
(\ref{eq:ots_mu1}) and (\ref{eq:ots_mu2}).

A sketch of the capacity region is shown with two upper bounds in
Fig.~\ref{fig:detailcap}(a). From Case 1 in Theorem
\ref{theorem:optdetail}, the point $B$ corresponds to the largest
transmission rate for the first user. The first upper bound is the
tangent of the capacity region at the point $B$, and its slope is
$-1/\varphi(\psi(1-\alpha_1))$. From Case 2, the point $A$ provides
the largest transmission rate for the second user. The second upper
bound is the tangent of the capacity region at the point $A$, and
its slope is $-1/\varphi(1)$. Case 3 gives us the optimal boundary
of the capacity region except the points $A$ and $B$.

Given $\alpha_1$ and $\alpha_2$, which completely describe a
two-user degraded broadcast Z channel, the optimal boundary of the
capacity region can be explicitly described by
(\ref{eq:ots_mu1}-\ref{eq:achI1}). For any $\mu_1$ in the range of
(\ref{eq:ots_mu1}), the value of the unique associated $\mu_2$
follows from (\ref{eq:ots_mu2}). The curve of the optimal boundary
of the capacity region is then the set of $(I_1,I_2)$ pairs
satisfying (\ref{eq:achI2}) and (\ref{eq:achI1}) for these $\mu_1$
and associated $\mu_2$. For example, for $\alpha_1=0.15$ and
$\alpha_2=0.6$, the range of optimal $\mu_1$ values is $0.445 \leq
\mu_1 \leq 1$, the range of optimal $\mu_2$ values implied by
(\ref{eq:ots_mu2}) is $0.392 \leq \mu_2 \leq 1$, and the associated
capacity region boundary is plotted in Fig.~\ref{fig:sim_rates}.

Now we prove Theorem \ref{theorem:strongopt}. Since we have proved
that the set of all the transmission strategies satisfying
(\ref{eq:ots_gamma}-\ref{eq:ots_mu2}) is sufficient, we only need to
show that any rate pair on the optimal boundary of the capacity
region can be achieved without using time sharing.

\emph{Proof by contradiction:} Suppose the point $Z$ in
Fig.~\ref{fig:detailcap}(b) is on the optimal boundary of the
capacity region for the broadcast Z channel and this point can only
be achieved by time sharing of the points $X$ and $Y$, which can be
directly achieved by using transmission strategies satisfying
(\ref{eq:ots_gamma}-\ref{eq:ots_mu2}). Clearly, the slope of the
line segment $XY$ is neither zero nor minus infinity. Denote
$-k,0<k<\infty$ as the slope of $XY$. The points $X$ and $Y$ provide
the same value of $ I_1+\frac{1}{k}I_2$. By Theorem
\ref{theorem:optdetail}, the optimal solution to the maximization
problem of $\max (I_1+\lambda I_2)$ is unique, and so neither $X$
nor $Y$ maximizes $(I_1+\frac{1}{k}I_2)$. Thus, there exists an
achievable point $P$ such that this point is on the right upper side
of the line $XY$. Since and the triangle $\triangle XYP$ is in the
capacity region, the point $Z$ must not be on the optimal boundary
of the capacity region (contradiction). Q.E.D.
\begin{figure}
  \centering
  \includegraphics[width=0.5\textwidth]{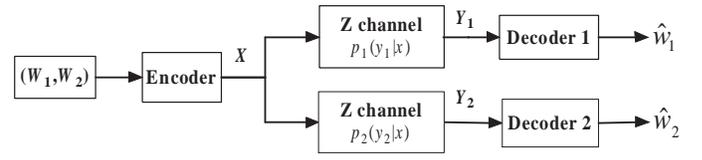}
  \caption{Communication system for 2-user broadcast Z channels.}\label{fig:ComSysBZC}\vspace{-0.1in}
\end{figure}
\begin{figure}
  \centering
  \includegraphics[width=0.5\textwidth]{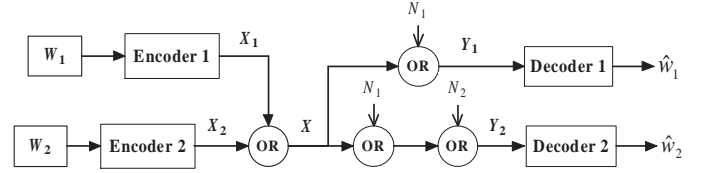}
  \caption{Optimal transmission strategy for broadcast Z channels.}\label{fig:OTSSysBZC}\vspace{-0.1in}
\end{figure}
\subsection{Independent encoding scheme}

The communication system for the two-user broadcast Z channel is
shown in Fig.~\ref{fig:ComSysBZC}. In a general scheme, the
transmitter jointly encodes the independent messages $W_1$ and
$W_2$, which is potentially too complex to implement. Theorem
\ref{theorem:strongopt} demonstrates that there exists an
independent encoding scheme which achieves the optimal boundary of
the capacity region. Since $\gamma =0$ is strongly sufficient, the
designed channel $X_2 \rightarrow X$ is a Z channel. Thus, the
broadcast signal $X$ can be constructed as the OR of two Bernoulli
random variables $X_1$ and $X_2$. This construction of $X$ is an
independent encoding scheme. The system diagram of the independent
encoding scheme is shown in Fig.~\ref{fig:OTSSysBZC}. First the
messages $W_1$ and $W_2$ are encoded separately and independently.
$X_1$ and $X_2$ are two binary random variables with
$\text{Pr}\{X_j=1\}=\bar{\mu}_j$ and $\text{Pr}\{X_j=0\}=\mu_j$,
where $\bar{\mu}_j+\mu_j=1$ for $j=1,2$. The transmitter broadcasts
$X$, which is the OR of $X_1$ and $X_2$. From Theorem
\ref{theorem:strongopt}, this independent encoding scheme with any
choice of $(\mu_1,\mu_2)$ satisfying (\ref{eq:ots_mu1}) and
(\ref{eq:ots_mu2}) achieves a rate pair $(I_1,I_2)$ arbitrarily
close to the optimal boundary of the capacity region if the codes
for $X_1$ and $X_2$ are properly chosen and have sufficiently large
block lengths.

\section{Nonlinear-Turbo Codes for the Two-User Broadcast Z Channel}
\label{sec:nlturbo}

\begin{figure}
\begin{center}
\scalebox{0.45} {\includegraphics{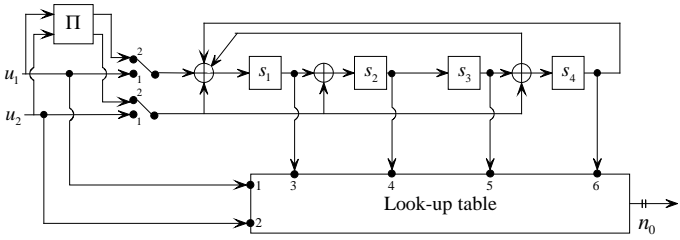}} \caption{16-state
nonlinear turbo code structure, with $k_0 = 2$ input bits per
trellis section.}
\label{fig:turbo_encoder}\end{center}\vspace{-0.2in}
\end{figure}

In this section we show a practical implementation of the
transmission strategy for the two-user broadcast Z channel. As
proved in Section \ref{sec:ots}, the optimal boundary is achieved by
transmitting the OR of the encoded data of each user, provided that
the density of ones of each of these encoded streams is chosen
properly. Hence, a family of codes that provides a controlled
density of ones is required. We use the nonlinear turbo codes,
introduced in \cite{griot-GC06}, to provide the needed controlled
density of ones. Nonlinear turbo codes are parallel concatenated
trellis codes with $k_0$ input bits and $n_0$ output bits per
trellis section. A look-up table assigns the output label for each
branch of the trellis so that the required ones density is achieved.
Each constituent encoder for the turbo code in this paper is a
16-state trellis code with $k_0 = 2$ and the trellis structure shown
in Fig.~\ref{fig:turbo_encoder}. The output labels are assigned via
a constrained search that provides the required ones density for
each case, using the tools presented in \cite{griot-GC06} for the Z
Channel. The output labels for the codes with rate pair
$(R_1=1/6,R_2=1/6)$, which is simulated on a broadcast Z channel
with $\alpha_1=0.15,\alpha_2=0.6$, are listed in Table
\ref{tab:NLTC-label}.

\begin{table}
\begin{center}
\caption{Labeling for constituent trellis codes. Rates $R_1=1/6,
R_2=1/6$. Rows represent the state $s_1 s_2 s_3 s_4$, columns
represent the input $u_1 u_2$. Labeling in octal
notation.}\label{tab:NLTC-label}
\begin{tabular}{|l||l|l|l|l|ll|l||l|l|l|l|}
\cline{1-5}\cline{8-12}
\multicolumn{5}{|c|}{User 1} & \multicolumn{1}{c}{} & \multicolumn{1}{c|}{} &  \multicolumn{5}{c|}{User 2} \\
\cline{1-5}\cline{8-12} state & \multicolumn{4}{c|}{input}&
\multicolumn{1}{c}{} & \multicolumn{1}{c|}{} &
state & \multicolumn{4}{c|}{input}\\
\cline{2-5}\cline{9-12}
& 00 & 01 & 10 & 11 &  & \multicolumn{1}{l|}{} &  & 00 & 01 & 10 & 11 \\
 \cline{1-5}\cline{8-12}
0000 & 40 & 20 & 10 & 04 &  &  & 0000 & 07 & 34 & 62 & 51 \\
\cline{1-5}\cline{8-12}
0001 & 20 & 40 & 04 & 10 &  &  & 0001 & 34 & 07 & 51 & 62 \\
\cline{1-5}\cline{8-12}
0010 & 10 & 04 & 02 & 01 &  &  & 0010 & 25 & 16 & 43 & 70 \\
\cline{1-5}\cline{8-12}
0011 & 04 & 10 & 01 & 02 &  &  & 0011 & 16 & 25 & 70 & 43 \\
\cline{1-5}\cline{8-12}
0100 & 02 & 01 & 40 & 20 &  &  & 0100 & 61 & 13 & 54 & 26 \\
\cline{1-5}\cline{8-12}
0101 & 01 & 02 & 20 & 40 &  &  & 0101 & 13 & 61 & 26 & 54 \\
\cline{1-5}\cline{8-12}
0110 & 42 & 21 & 14 & 05 &  &  & 0110 & 23 & 15 & 52 & 64 \\
\cline{1-5}\cline{8-12}
0111 & 21 & 42 & 05 & 14 &  &  & 0111 & 15 & 23 & 64 & 52 \\
\cline{1-5}\cline{8-12}
1000 & 01 & 02 & 04 & 10 &  &  & 1000 & 70 & 43 & 16 & 25 \\
\cline{1-5}\cline{8-12}
1001 & 02 & 01 & 10 & 04 &  &  & 1001 & 43 & 70 & 25 & 16 \\
\cline{1-5}\cline{8-12}
1010 & 04 & 10 & 20 & 40 &  &  & 1010 & 51 & 62 & 34 & 07 \\
\cline{1-5}\cline{8-12}
1011 & 10 & 04 & 40 & 20 &  &  & 1011 & 62 & 51 & 07 & 34 \\
\cline{1-5}\cline{8-12}
1100 & 05 & 14 & 21 & 42 &  &  & 1100 & 64 & 52 & 15 & 23 \\
\cline{1-5}\cline{8-12}
1101 & 14 & 05 & 42 & 21 &  &  & 1101 & 52 & 64 & 23 & 15 \\
\cline{1-5}\cline{8-12}
1110 & 20 & 40 & 01 & 02 &  &  & 1110 & 26 & 54 & 13 & 61 \\
\cline{1-5}\cline{8-12}
1111 & 40 & 20 & 02 & 01 &  &  & 1111 & 54 & 26 & 61 & 13 \\
\cline{1-5}\cline{8-12}
\end{tabular}

\end{center}
\end{table}

\begin{figure}
\begin{center}
\scalebox{0.45} {\includegraphics{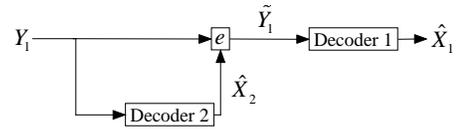}} \caption{Decoder
structure for user 1.}
\label{fig:decoder_user1}\end{center}\vspace{-0.2in}
\end{figure}

Receiver 1 uses successive decoding as shown in
Fig.~\ref{fig:decoder_user1}. Denote as $\hat{X}_2$ the decoded
stream corresponding to user 2. Since the transmitted data is $x =
x_1 (\text{OR}) x_2$, whenever a bit $x_2 = 1$, there is no
information about $x_1$, and $x_1$ can be considered an erasure.
Hence, the input stream to Decoder 1 is
\begin{eqnarray}
\hat{y}_1=e(y_1,\hat{x}_2)= \left\{\begin{array}{r c l}
y_1 & \text{if} & \hat{x}_2=0, \\
e & \text{if} & \hat{x}_2=1.
\end{array}\right.
\end{eqnarray}

\begin{figure}
\begin{center}
\scalebox{0.45} {\includegraphics{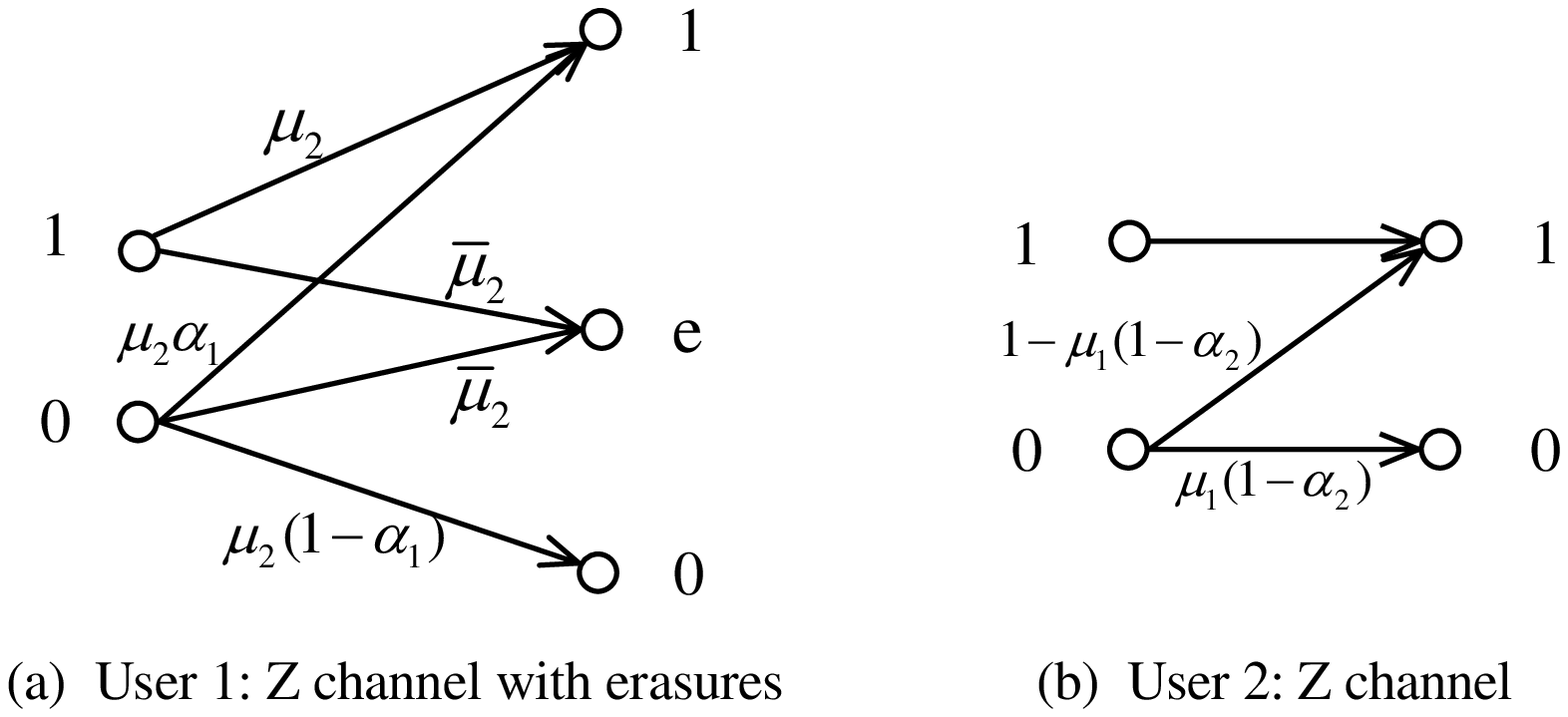}} \caption{Perceived
channel by each decoder.}
\label{fig:channels}\end{center}\vspace{-0.2in}
\end{figure}

Therefore, Decoder 2 sees a Z Channel with erasures as shown in
Fig.~\ref{fig:channels}. The tools presented in \cite{griot-GC06}
were general enough to be applied to the Z Channel with erasures.
Note that if $\alpha_1$ is much smaller than $\alpha_2$ we can use
hard decoding in Decoder 2 instead of soft decoding without any loss
in performance. Since the code for user 2 is designed for a Z
Channel with 0-to-1 crossover probability $1-(1-\alpha_2)\mu_1$, and
the channel perceived by Decoder 2 in user 1 is a Z-Channel with
crossover probability $1 - (1-\alpha_1)\mu_1 < 1-(1-\alpha_2)\mu_1$,
the bit error rate of $\hat{x}_2$ is negligible compared to the bit
error rate of Decoder 1. In fact, in all the simulations shown in
Section \ref{sec:results}, which include 100 frame errors of user 1,
none of the errors were produced by Decoder 2.

\section{Results}
\label{sec:results}

\begin{figure}
\begin{center}
\scalebox{0.6} {\includegraphics{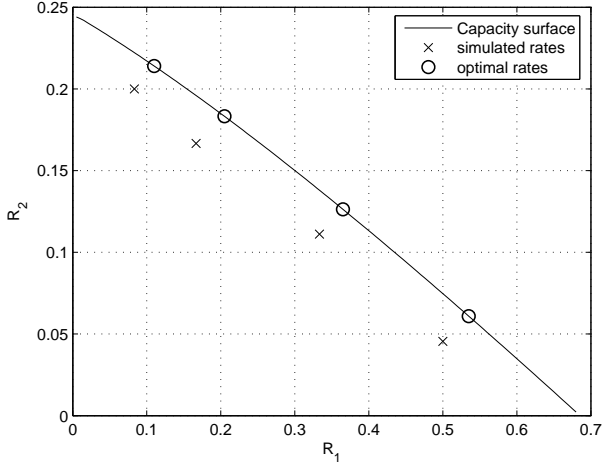}} \caption{Broadcast Z
channel with crossover probabilities $\alpha_1 = 0.15$ and $\alpha_2
= 0.6$ for receiver 1 and 2 respectively: achievable capacity
region, simulated rate pairs ($R_1,R_2$) and their corresponding
optimal rates.} \label{fig:sim_rates}\end{center}\vspace{-0.2in}
\end{figure}

We simulate the transmission strategy for the two-user broadcast Z
channel with crossover probabilities $\alpha_1 = 0.15$ and $\alpha_2
= 0.6$, using nonlinear turbo codes, with the structure shown in
Fig.~\ref{fig:turbo_encoder}. Fig.~\ref{fig:sim_rates} shows the
capacity region for the broadcast Z channel and identifies the
simulated rate pairs. It also shows the optimal rate pairs, which
are used to compute the ones densities of each code. The output
labels for the codes with each simulated rate pair are listed at
\cite{CSLcode}. For each of these four simulated rate pairs, the
loss in mutual information from the associated optimal rate is only
0.04 bits or less in $R_1$ and only 0.02 bits or less in $R_2$.
Table \ref{tab:NLTC-or-mac} shows bit error rates for each rate
pair, the ones densities $\bar{\mu}_1$ and $\bar{\mu}_2$, and the
interleaver lengths $K_1$ and $K_2$ used for each code. For
simplicity, we chose $K_1$ and $K_2$ so that the codeword length $n$
would be the same for user 1 and user 2, except for rate pairs $R_1
= 1/2$ and $R_2 = 1/22$, where one codeword length of user 2 is
twice the length of user 1.

\begin{table*}
\begin{center}
\caption{BER for two-user broadcast Z channel with crossover
probabilities $\alpha_1 = 0.15$ and $\alpha_2 =
0.6$.\label{tab:NLTC-or-mac}}
\begin{tabular}{|c|c|c|c|c|c|c|c|}
\hline $R_1$ &  $R_2$  &  $\bar{\mu}_1$ & $\bar{\mu}_2$ & $K_1$ & $K_2$ & BER${_1}$ & BER$_2$\\
\hline $1/12$ & $1/5$  & 0.106 & 0.56 & 4800 & 1700 & $2.54\times10^{-5}$ & $1.24\times10^{-5}$ \\
\hline $1/6$ & $1/6$ & 0.196 & 0.5 & 2048  & 2048 & $7.01\times10^{-6}$ & $5.33\times10^{-6}$\\
\hline $1/3$ & $1/9$ & 0.336 & 0.3739 & 4608 & 1536 & $7.13\times10^{-6}$ & $6.70\times10^{-6}$\\
\hline $1/2$ & $1/22$ & 0.463 & 0.1979 & 5632 & 1024 & $9.27\times10^{-7}$ & $3.27\times10^{-6}$\\
\hline
\end{tabular}
\end{center}
\end{table*}

\section{Conclusions}
\label{sec:conclusions}

This paper presented an optimal transmission strategy for the
broadcast Z channel with independent encoding and successive
decoding. We proved that any point on the optimal boundary of the
capacity region can be achieved by independently encoding the
messages corresponding to different users and transmitting the OR of
the encoded signals. Also, the distributions of the outputs of each
encoder that achieve the optimal boundary were provided.
Nonlinear-turbo codes that provide a controlled distribution of ones
and zeros in their codewords were used to demonstrate a
low-complexity scheme that works close to the optimal boundary.

\section*{Appendices}
\subsection*{Appendix A}\label{app:A}
Here we prove Theorem \ref{theorem:notopt}, which states that for a
broadcast Z channel with $0<\alpha_1<\alpha_2<1$, any transmission
strategy $(\mu_1,\mu_2,\gamma)$ with $0< \mu_2<1,0< \gamma< \mu_1$
is not optimal.

In (\ref{eq:capI2}) and (\ref{eq:capI1}), denote
\begin{align}
I_1(\mu_1,\mu_2,\gamma)&=I(X;Y_1|X_2)\big| _{\mu_1,\mu_2,\gamma}\\
I_2(\mu_1,\mu_2,\gamma)&=I(X_2;Y_2)\big| _{\mu_1,\mu_2,\gamma}\\
I_{1,2}(\mu_1,\mu_2,\gamma)&=(I_1,I_2)\big| _{\mu_1,\mu_2,\gamma}.
\end{align}
The transmission strategy $(\mu_1,\mu_2,\gamma)$ achieves the rate
pair $I_{1,2}(\mu_1,\mu_2,\gamma)$. The theorem is true if we can
increase both $I_1$ and $I_2$ when $0<\mu_2<1,0<\gamma<\mu_1$.

First compare the strategies $(\mu_1,\mu_2,\gamma)$ and
$(\mu_1+\bar{\mu}_2\delta_1,\mu_2,\gamma-\mu_2\delta_1)$ for a small
positive number $\delta_1>0$.
\begin{align}\label{eq:d1I1}
\Delta_1 I_1 & = I_1(\mu_1+\bar{\mu}_2\delta_1,\mu_2,\gamma-\mu_2\delta_1)-I_1(\mu_1,\mu_2,\gamma){}\nonumber\\
& \simeq
\frac{\partial I_1(\mu_1+\bar{\mu}_2\delta_1,\mu_2,\gamma-\mu_2\delta_1)}{\partial\delta_1}\Big| _{\delta_1=0}\delta_1{}\nonumber\\
& = -\mu_2\bar{\mu}_2(1-\alpha_1) \Big \{ \ln
\frac{1-\gamma(1-\alpha_1)}{\gamma(1-\alpha_1)} \nonumber \\
& \qquad \qquad \qquad + \ln
\frac{\mu_1(1-\alpha_1)}{1-\mu_1(1-\alpha_1)} \Big \}\delta_1\nonumber\\
& < 0,
\end{align}
and
\begin{align}\label{eq:d1I2}
\Delta_1 I_2 & = I_2(\mu_1+\bar{\mu}_2\delta_1,\mu_2,\gamma-\mu_2\delta_1)-I_2(\mu_1,\mu_2,\gamma){}\nonumber\\
& \simeq
\frac{\partial I_2(\mu_1+\bar{\mu}_2\delta_1,\mu_2,\gamma-\mu_2\delta_1)}{\partial\delta_1}\Big| _{\delta_1=0}\delta_1{}\nonumber\\
& = \mu_2\bar{\mu}_2(1-\alpha_2)\Big\{\ln
\frac{1-\gamma(1-\alpha_2)}{\gamma(1-\alpha_2)} \nonumber \\
& \qquad \qquad \qquad +\ln
\frac{\mu_1(1-\alpha_2)}{1-\mu_1(1-\alpha_2)}\Big\}\delta_1 \nonumber\\
& >0.
\end{align}

\begin{figure}
  \centering
  \includegraphics[width=0.3\textwidth]{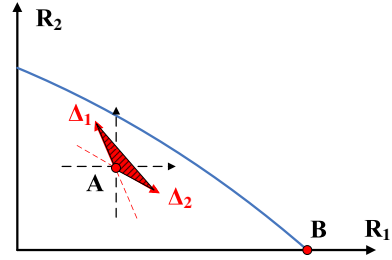}
  \caption{Capacity region and the changes of rate pairs.}\label{fig:ratepairmove}\vspace{-0.1in}
\end{figure}

The small change of the rate pair $(\Delta_1I_1,\Delta_1I_2)$ is
shown Fig.~\ref{fig:ratepairmove}. Point $A$ is the rate pair of the
transmission strategy $(\mu_1,\mu_2,\gamma)$, the arrow $\Delta_1$
shows the small movement of the rate pair
$(\Delta_1I_1,\Delta_1I_2)$.

Second compare the strategies $(\mu_1,\mu_2,\gamma)$ and
$(\mu_1+(\gamma-\mu_1)\delta_2,\mu_2+\mu_2\delta_2,\gamma)$ for a
small positive number $\delta_2>0$.
\begin{align}\label{eq:d2I1}
\Delta_2 I_1 & = I_1(\mu_1+(\gamma-\mu_1)\delta_2,\mu_2+\mu_2\delta_2,\gamma)-I_1(\mu_1,\mu_2,\gamma){}\nonumber\\
& \simeq
\frac{\partial I_1(\mu_1+(\gamma-\mu_1)\delta_2,\mu_2+\mu_2\delta_2,\gamma)}{\partial\delta_2}\Big|_{\delta_2=0}\delta_2{}\nonumber\\
& = - \mu_2\delta_2\Big\{\gamma(1-\alpha_1)\ln \frac{\mu_1}{\gamma} \nonumber \\
& \qquad \qquad + (1-\gamma(1-\alpha_1))\ln \frac{1-\mu_1(1-\alpha_1)}{1-\gamma(1-\alpha_1)}\Big\}{}\nonumber\\
& = \mu_2\delta_2D(\gamma(1-\alpha_1)\parallel \mu_1(1-\alpha_1))\nonumber \\
& > 0,
\end{align}
and
\begin{align}\label{eq:d2I2}
\Delta_2 I_2 & = I_2(\mu_1+(\gamma-\mu_1)\delta_2,\mu_2+\mu_2\delta_2,\gamma)-I_2(\mu_1,\mu_2,\gamma){}\nonumber\\
& \simeq
\frac{\partial I_2(\mu_1+(\gamma-\mu_1)\delta_2,\mu_2+\mu_2\delta_2,\gamma)}{\partial\delta_2}\Big|_{\delta_2=0}\delta_2{}\nonumber\\
& =  \mu_2\delta_2\Big\{\gamma(1-\alpha_2)\ln \frac{\mu_1}{\gamma} \nonumber \\
& \qquad \quad +(1-\gamma(1-\alpha_2))\ln \frac{1-\mu_1(1-\alpha_2)}{1-\gamma(1-\alpha_2)}\Big\}{}\nonumber\\
& =  - \mu_2\delta_2D(\gamma(1-\alpha_2)\parallel \mu_1(1-\alpha_2))\nonumber \\
& < 0,
\end{align}
where $D(p \parallel q)$ is the relative entropy between the
distributions $p$ and $q$. The arrow $\Delta_2$ in
Fig.~\ref{fig:ratepairmove} shows the small movement of the rate
pair $(\Delta_2I_1,\Delta_2I_2)$.

Now we show that
\begin{equation}\label{eq:slope}
\frac{\Delta_1I_2}{\Delta_1I_1}<\frac{\Delta_2I_2}{\Delta_2I_1}<0.
\end{equation}

\begin{align}
& \quad \frac{\Delta_1I_2}{\Delta_1I_1}<\frac{\Delta_2I_2}{\Delta_2I_1}{}\nonumber\\
& \Leftrightarrow  \frac{D(\gamma(1-\alpha_2)\parallel
\mu_1(1-\alpha_2))+\ln
\frac{1-\gamma(1-\alpha_2)}{1-\mu_1(1-\alpha_2)}}{D(\gamma(1-\alpha_1)\parallel
\mu_1(1-\alpha_1))+\ln
\frac{1-\gamma(1-\alpha_1)}{1-\mu_1(1-\alpha_1)}} \nonumber \\
& \qquad \qquad > \frac{D(\gamma(1-\alpha_2)\parallel
\mu_1(1-\alpha_2))}{D(\gamma(1-\alpha_1)\parallel
\mu_1(1-\alpha_1))}{}\nonumber\\
& \Leftrightarrow  \!\! \frac{D(\gamma(1-\!\alpha_1)\!\parallel \!
\mu_1(1-\!\alpha_1))}{\ln
\frac{1-\gamma(1-\alpha_1)}{1-\mu_1(1-\alpha_1)}} \!
> \!\frac{D(\gamma(1-\! \alpha_2)\! \parallel \!\mu_1(1-\! \alpha_2))}{\ln
\frac{1-\gamma(1-\alpha_2)}{1-\mu_1(1-\alpha_2)}}{}\nonumber\\
& \Leftrightarrow  \!f(x) \!=\!\frac{D(\gamma x \! \parallel \!
\mu_1x)}{\ln \frac{1-\gamma x}{1-\mu_1x}} \textrm{\small{is
monotonically
increasing in}} \quad \!\!\! 0\!<\!x\!<\!1 {}\nonumber\\
& \Leftrightarrow f'(x)=\Big \{\ln\frac{\gamma
x}{\mu_1x}\ln\frac{1-\gamma x}{1-\mu_1x}-\big (\ln\frac{1-\gamma
x}{1-\mu_1x} \big )^2 \nonumber \\
& \qquad \qquad \! +\ln\frac{\gamma x}{\mu_1x}\big(\frac{1}{1-\gamma
x}-\frac{1}{1-\mu_1x}\big)\Big\}\gamma \big(\ln \frac{1-\gamma
x}{1-\mu_1x} \big )^{-2} \nonumber \\
& \quad \quad \quad \quad >0.
\end{align}

Let $a=1-\gamma x$ and $b=1-\mu_1x$. We have $0<b<a<\!1$ and want to
show that
\begin{equation}
g(a,b) = \ln \frac{a}{b} \ln \frac{1-a}{1-b} -\! \left( \ln
\frac{a}{b} \right )^2 \!\!+ \ln \frac{1-a}{1-b} \left (\frac{1}{a}
- \frac{1}{b} \right )>0.
\end{equation}
Since
\begin{equation}
\frac{\partial^2 g(a,b)}{\partial a\partial
b}=-\frac{(a-b)^2}{a^2b^2(1-a)(1-b)}<0,
\end{equation}
and
\begin{equation}
\frac{\partial g(a,b)}{\partial a}\Big|_{b=a}=0 \quad \forall 0<a<1,
\end{equation}
it is true that
\begin{equation}\label{eq:guv}
\frac{\partial g(a,b)}{\partial a}>0 \quad \forall 0<b<a<1.
\end{equation}
It follows from (\ref{eq:guv}) and the fact $g(b,b)=0, \forall
0<b<1$ that $g(a,b)>0,\forall 0<b<a<1$. Thus, the inequality
(\ref{eq:slope}) is true, which means that the slope of $\Delta_1$
is smaller than that of $\Delta_2$ in Fig.~\ref{fig:ratepairmove}.
Hence, the achievable shaded region is on the upper right side of
the point $A$. Therefore, we can increase both terms in the rate
pair $I_{1,2}(\mu_1,\mu_2,\gamma)$ simultaneously and the strategy
$(\mu_1,\mu_2,\gamma)$ is not optimal when $0<\mu_2<1$ and $0<\gamma
<\mu_1$. Q.E.D.

\subsection*{Appendix B}\label{app:B}
Here we prove Theorem \ref{theorem:optdetail}, which provides the
unique optimal solution to the maximization problem (\ref{eq:max}).
In problem (\ref{eq:max}), the objective function $I_1+\lambda I_2$
is bounded and the domain $0\leq \mu_1,\mu_2\leq1$ is closed, so the
maximum exists and can be attained. First we discuss some possible
optimal solutions and then we show that only one of them is optimal for any fixed $\lambda \geq 0$. \\
\emph{Case 0:} If $\mu_1=0$ or $\mu_2=0$ or $\mu_1=\mu_2=1$, then
$I_1=I_2=0$ and
so it cannot be optimal.\\
\emph{Case 1:} If $\mu_2=1$ and $0<\mu_1<1$, then $I_2=0$ and
\begin{equation}
\frac{\partial I_1}{\partial \mu_1}=(1-\alpha_1)\ln\frac{1-\mu_1
(1-\alpha_1)}{\mu_1 (1-\alpha_1)}-H(1-\alpha_1)=0
\end{equation}
\begin{equation}
\Rightarrow
\mu_1^*=\frac{1}{(1-\alpha_1)(e^{H(1-\alpha_1)/(1-\alpha_1)}+1)}.
\end{equation}
\emph{Case 2:} If $\mu_1=1$ and $0<\mu_2<1$, then $I_1=0$ and
\begin{equation}
\frac{\partial I_2}{\partial \mu_2}=(1-\alpha_2)\ln\frac{1-\mu_2
(1-\alpha_2)}{\mu_2 (1-\alpha_2)}-H(1-\alpha_2)=0
\end{equation}
\begin{equation}
\Rightarrow
\mu_2^*=\frac{1}{(1-\alpha_2)(e^{H(1-\alpha_2)/(1-\alpha_2)}+1)}.
\end{equation}
\emph{Case 3:} If $0<\mu_1,\mu_2<1$, then the optimum is attained
when
\begin{equation}
\mu_2\frac{\partial (I_1+\lambda I_2)}{\partial \mu_2} -
\mu_1\frac{\partial ( I_1+\lambda I_2)}{\partial \mu_1}=0\nonumber
\end{equation}
\begin{equation}\label{eq:betaeqa}
\Rightarrow \ln(1-\mu_1^*(1-\alpha_1)) = \lambda
\ln(1-\mu_1^*(1-\alpha_2)),
\end{equation}
and
\begin{align}\label{eq:alphaeqa}
& \frac{\partial (I_1+\lambda I_2)}{\partial
\mu_2}=0{}\nonumber\\
\Rightarrow &\lambda \left \{H(\mu_1^*(1-\alpha_2))-
\mu_1^*(1-\alpha_2)\ln\frac{1-\mu_2^*\mu_1^*(1-\alpha_2)}{\mu_2^*\mu_1^*(1-\alpha_2)}\right \}\nonumber \\
& =\big(H(\mu_1^*(1-\alpha_1))-\mu_1^*H(1-\alpha_1)\big){}\nonumber\\
\Rightarrow & \big(H(\mu_1^*(1-\alpha_1))-\mu_1^*H(1-\alpha_1)\big)
\cdot \ln(1-\mu_1^*(1-\alpha_2))\nonumber \\
& = \left \{H(\mu_1^*(1-\alpha_2)) -
\mu_1^*(1-\alpha_2)\ln\frac{1-\mu_2^*\mu_1^*(1-\alpha_2)}{\mu_2^*\mu_1^*(1-\alpha_2)}\right\} \nonumber\\
& \quad \cdot \ln(1-\mu_1^*(1-\alpha_1)).
\end{align}

For any fixed $\lambda \geq 0$, the optimal solution is in Case 1,
2, or 3.

\newtheorem{lemma1}{Lemma}
\begin{lemma1}\label{theorem:lemma1}
Function
$\varphi(x)=\frac{\ln(1-(1-\alpha_1)x)}{\ln(1-(1-\alpha_2)x)}$ is
monotonically increasing in the domain of $0\leq x\leq1$ when
$\alpha_1<\alpha_2$.
\end{lemma1}

\newtheorem{notcase1}[lemma1]{Lemma}
\begin{notcase1}\label{theorem:notcase1}
The solution in Case 1 cannot be optimal when
$\lambda>\varphi(\psi(1-\alpha_1))$.
\end{notcase1}
\emph{Proof:} When $\mu_2=1$ and $\mu_1=\psi(1-\alpha_1)$,
$\frac{\partial I_2}{\partial \mu_1}=0$ and $\frac{\partial
I_1}{\partial \mu_1}=0$. Therefore, for any fixed $\lambda$,
$\frac{\partial (I_1+\lambda I_2)}{\partial \mu_1}=0$. When
$\lambda=\varphi(\mu_1)=\varphi(\psi(1-\alpha_1))$,
(\ref{eq:betaeqa}) holds, and so
\begin{align}
& \quad \frac{\partial (I_1+\lambda I_2)}{\partial \mu_2}\Big|
_{\mu_2=1,\mu_1=\psi(1-\alpha_1)} \nonumber \\
& = \frac{\partial (I_1+\varphi(\psi(1-\alpha_1)) I_2)}{\partial
\mu_2}\Big| _{\mu_2=1,\mu_1=\psi(1-\alpha_1)}\nonumber \\
& =0. \label{eq:notcase1}
\end{align}
When $\lambda>\varphi(\psi(1-\alpha_1))$,
\begin{align}
&\frac{\partial (I_1+\lambda I_2)}{\partial \mu_2}\Big| _{\mu_2=1,\mu_1=\psi(1-\alpha_1)}\nonumber\\
&=\frac{\partial I_1}{\partial \mu_2}\Big|
_{\mu_2=1,\mu_1=\psi(1-\alpha_1)}+\lambda\frac{\partial
I_2}{\partial
\mu_2}\Big| _{\mu_2=1,\mu_1=\psi(1-\alpha_1)}\nonumber\\
& \overset{a}{<}  \frac{\partial I_1}{\partial \mu_2}\Big|
_{\mu_2=1,\mu_1=\psi(1-\alpha_1)}\!\!\!\!\!\!\!\!\!+\varphi(\psi(1-\alpha_1))
\frac{\partial I_2}{\partial
\mu_2}\Big| _{\mu_2=1,\mu_1=\psi(1-\alpha_1)}\nonumber\\
& = \frac{\partial (I_1+\varphi(\psi(1-\alpha_1)) I_2)}{\partial
\mu_2} \Big| _{\mu_2=1,\mu_1=\psi(1-\alpha_1)} \nonumber \\
& \overset{b}{=} 0,
\end{align}
where (a) follows from the facts that $\frac{\partial I_2}{\partial
\mu_2}\Big|_{\mu_2=1,\mu_1=\psi(1-\alpha_1)}=\ln(1-\psi(1-\alpha_1)\cdot(1-\alpha_2))<0$
and $\lambda>\varphi(\psi(1-\alpha_1))$, and (b) follows from
(\ref{eq:notcase1}). Therefore, Case 1 cannot be optimal when
$\lambda>\varphi(\psi(1-\alpha_1))$. Q.E.D.

\newtheorem{notcase2}[lemma1]{Lemma}
\begin{notcase2}\label{theorem:notcase2}
The solution in Case 2 cannot be optimal when $\lambda<\varphi(1)$.
\end{notcase2}
\emph{Proof:} When $\mu_2=\psi(1-\alpha_2)$ and $\mu_1=1$,
$\frac{\partial I_2}{\partial \mu_2}=0$ and $\frac{\partial
I_1}{\partial \mu_2}=0$. Therefore, for any fixed $\lambda$,
$\frac{\partial (I_1+\lambda I_2)}{\partial \mu_2}=0$. When
$\lambda=\varphi(\mu_1)=\varphi(1)$, (\ref{eq:betaeqa}) holds, and
so
\begin{align}
& \quad \frac{\partial (I_1+\lambda I_2)}{\partial \mu_1}\Big|
_{\mu_2=\psi(1-\alpha_2),\mu_1=1} \nonumber\\
& = \frac{\partial (I_1+\varphi(1) I_2)}{\partial \mu_1} \Big|
_{\mu_2=\psi(1-\alpha_2),\mu_1=1} \nonumber \\
& =0. \label{eq:notcase2}
\end{align}
When $\lambda<\varphi(1)$,
\begin{align}
&\frac{\partial (I_1+\lambda I_2)}{\partial \mu_1}\Big| _{\mu_2=\psi(1-\alpha_2),\mu_1=1}\nonumber\\
&=\frac{\partial I_1}{\partial \mu_2}\Big|
_{\mu_2=\psi(1-\alpha_2),\mu_1=1}+\lambda\frac{\partial
I_2}{\partial
\mu_2}\Big| _{\mu_2=\psi(1-\alpha_2),\mu_1=1}\nonumber\\
& \overset{a}{<} \frac{\partial I_1}{\partial \mu_2}\Big|
_{\mu_2=\psi(1-\alpha_2),\mu_1=1}+\varphi(1) \frac{\partial
I_2}{\partial \mu_2}\Big| _{\mu_2=\psi(1-\alpha_2),\mu_1=1}\nonumber\\
& = \frac{\partial (I_1+\varphi(1) I_2)}{\partial \mu_1} \Big|
_{\mu_2=\psi(1-\alpha_2),\mu_1=1} \nonumber \\
& \overset{b}{=} 0,
\end{align}
where (a) follows from the facts that $\frac{\partial I_2}{\partial
\mu_1}\Big|_{\mu_2=\psi(1-\alpha_2),\mu_1=1}= -\psi (1- \alpha_2)
\ln \alpha_2>0$ and $\lambda<\varphi(1)$, and (b) follows from
(\ref{eq:notcase2}). Therefore, Case 2 cannot be optimal when
$\lambda<\varphi(1)$. Q.E.D.

\newtheorem{lemma2}[lemma1]{Lemma}
\begin{lemma2}\label{theorem:lemma2}
The solution to (\ref{eq:betaeqa}) exists in $(0,1)$ and is unique
for any $\lambda$ in the range of $\varphi(0)<\lambda<\varphi(1)$.
\end{lemma2}

\emph{Proof:} Equation (\ref{eq:betaeqa}) is equivalent to
$\varphi(\mu_1^*)=\lambda$. From Lemma \ref{theorem:lemma1},
 $\varphi(\mu_1)$ is monotonically increasing. Therefore, when $\varphi(0)<\lambda<\varphi(1)$, the solution $\mu_1^*$is unique and
$\mu_1^*\in(0,1)$. Q.E.D.

\newtheorem{lemma3}[lemma1]{Lemma}
\begin{lemma3}\label{theorem:lemma3}
The unique solution $(\mu_1^*,\mu_2^*)$ to (\ref{eq:betaeqa}) and
(\ref{eq:alphaeqa}) in Case 3 is optimal if
$\varphi(\psi(1-\alpha_1))<\lambda<\varphi(1)$.
\end{lemma3}

\emph{Proof:} From Lemma \ref{theorem:lemma2}, the solution
$\mu_1^*$ to (\ref{eq:betaeqa}) is unique if
$\varphi(\psi(1-\alpha_1))<\lambda<\varphi(1)$. From
(\ref{eq:alphaeqa}),
\begin{align}
& \qquad m(\mu_2) \nonumber\\
&=\Big\{H(\mu_1^*(1-\alpha_2))-\mu_1^*(1-\alpha_2)\ln\frac{1-\mu_2\mu_1^*(1-\alpha_2)}{\mu_2\mu_1^*(1-\alpha_2)}\Big\}
\nonumber \\
&  \quad \! \cdot \ln(1-\mu_1^*(1-\alpha_1))\!-\!\Big\{H(\mu_1^*(1-\alpha_1))-\mu_1^*H(1-\alpha_1)\Big\} \nonumber \\
& \quad \cdot \ln(1-\mu_1^*(1-\alpha_2)){}\nonumber\\
&=0.
\end{align}
Clearly, $m(\mu_2)$ is monotonically increasing,
\begin{equation}
\lim_{\mu_2\rightarrow0}m(\mu_2)=-\infty<0,
\end{equation}
and
\begin{align}
&\varphi(\psi(1-\alpha_1))<\lambda<\varphi(1)\nonumber\\
\Rightarrow & \mu_1^*>\psi(1-\alpha_1) \nonumber\\
\Rightarrow & m(1)>0.
\end{align}
That means the unique solution $\mu_2^*$ to (\ref{eq:alphaeqa}) is
in the domain of $0\leq \mu_2\leq1$. Furthermore, when
$\varphi(\psi(1-\alpha_1))<\lambda<\varphi(1)$, by Lemma
\ref{theorem:notcase1} and Lemma \ref{theorem:notcase2}, Case 1 or
Case 2 cannot be optimal because
\begin{equation}\label{eq:case1}
\frac{\partial( I_1+ \lambda I_2)}{\partial
\mu_2}\big|_{\mu_2=1,\mu_1=\psi(1-\alpha_1)}<0,
\end{equation}
\begin{equation}\label{eq:case2}
\frac{\partial( I_1+ \lambda I_2)}{\partial
\mu_1}\big|_{\mu_1=1,\mu_2=\psi(1-\alpha_2)}<0.
\end{equation}
Therefore, Case 3 is optimal. Q.E.D.

\newtheorem{lemma4}[lemma1]{Lemma}
\begin{lemma4}\label{theorem:lemma4}
The unique solution $(\mu_2^*=1,\mu_1^*=\psi(1-\alpha_1))$ in Case 1
is optimal if $0 \leq\lambda\leq \varphi(\psi(1-\alpha_1))$.
\end{lemma4}

\emph{Proof:} When $0 \leq\lambda\leq \varphi(\psi(1-\alpha_1))$,
Case 3 is not optimal because there is no solution $\mu_1\in(0,1)$
to (\ref{eq:betaeqa}). Case 2 is not optimal by Lemma
\ref{theorem:notcase2}. Hence, Case 1 is optimal. Q.E.D.

\newtheorem{lemma5}[lemma1]{Lemma}
\begin{lemma5}\label{theorem:lemma5}
The unique solution $(\mu_2^*=\psi(1-\alpha_2),\mu_1^*=1)$ in Case 2
is optimal if $\lambda\geq\varphi(1)$.
\end{lemma5}

\emph{Proof:} When $\lambda\geq\varphi(1)$, Case 3 is not optimal
because there is no solution $\mu_2\in(0,1)$ to (\ref{eq:alphaeqa}).
Case 1 is not optimal by Lemma \ref{theorem:notcase1}. Hence, Case 2
is optimal. Q.E.D.

From Lemma \ref{theorem:lemma3},\ref{theorem:lemma4} and
\ref{theorem:lemma5}, Theorem \ref{theorem:optdetail} is immediately
proved. Q.E.D.


\begin{biography}{Bike Xie}
(S'07) was born in Shanghai, China, in 1983. He received the B.S.
degree in electronic engineering from Tsinghua University, Beijing,
China, in 2005, and the M.S. degree in electrical engineering from
University of California, Los Angeles, CA, in 2006. He is currently
working toward the Ph.D. degree in the Communication Systems Lab at
the Department of Electrical Engineering, University of California,
Los Angeles. \\
His research interests are in the area of information
theory with particular interest in the topics of network information
theory and channel coding.
\end{biography}

\begin{biography}{Miguel Griot}
(S'05) received the B.S. degree in electrical engineering from the
Universidad de la Republica, Uruguay, in 2003, the M.S. and PhD
degree in electrical engineering from the University of California
at Los Angeles, in 2004 and 2008 respectively. He is currently with
Qualcomm Inc., San Diego. His research interests include wireless
communications, channel coding, information theory, multiple access
channels and broadcast channels.
\end{biography}

\begin{biography}{Andres I. Vila Casado}
received his B. S. in electrical engineering from the Politecnico di
Torino, Turin, Italy in 2002. He received his M. S. and Ph. D.
degrees in electrical engineering from the University of California,
Los Angeles in 2004 and 2007 respectively. \\
At UCLA and at Politecnico di Torino he conducted research on
communication theory with a focus on channel coding and information
theory. He is currently a Research Scientist at Mojix, Inc. where he
conducts research on physical layer communications and Bayesian
estimation for RFID applications.
\end{biography}

\begin{biography}{Richard D. Wesel}
is a Professor with the UCLA Electrical Engineering Department and
is the Associate Dean for Academic and Student Affairs for the UCLA
Henry Samueli School of Engineering and Applied Science. He joined
UCLA in 1996 after receiving his Ph.D. in electrical engineering
from Stanford. His B.S. and M.S. degrees in electrical engineering
are from MIT.  His research is in the area of communication theory
with particular interest in channel coding. He has received the
National Science Foundation (NSF) CAREER Award, an Okawa Foundation
award for research in information and telecommunications, and the
Excellence in Teaching Award from the Henry Samueli School of
Engineering and Applied Science.  He has authored or co-authored
over a hundred conference and journal publications.
\end{biography}
\newpage

\end{document}